\documentclass[default,iicol]{sn-jnl}        % lineno shows the line numbering

\usepackage[utf8]{inputenc}
\usepackage{lmodern}                         % A lot of warnings arose due to the misuse of fonts, so this is added
\usepackage{fix-cm}                          % A lot of warnings arose due to the misuse of fonts, so this is added
\usepackage{placeins}                        % Add this in the preamble
\usepackage{upgreek}                         % Added to create upright mu -> $\mathrm{\upmu}$
\usepackage[version=4]{mhchem}               % Added to create decent chemical formulas -> \ce{Si3N4}
\usepackage{titlesec}                        % Changing the Appendix header to S.I.
\usepackage{hyperref}                        % For better URL breaking and hyperlinks
\usepackage{array}

\jyear{2025}

\theoremstyle{thmstyleone}
\theoremstyle{thmstyletwo}

\theoremstyle{thmstylethree}

\titleformat{\section}
  {\normalfont\large\bfseries\raggedright}  % formatting
  {S.I.~\thesection}                        % label
  {1em}                                     % spacing between label and title
  {}                                        % code before title
\titlespacing*{\section}{0pt}{*3}{*1}

\usepackage{graphics}

\DeclareUnicodeCharacter{2212}{-}

\newcolumntype{L}[1]{>{\raggedright\arraybackslash}p{#1}}

\DeclareMathOperator{\sech}{sech}

\begin{document}

\title[Article Title]{Stable magnetic nanodomains engineered via Ga$^{+}$-ion irradiation for deterministic sequential switching}

%%=======================================================%%
%% Prefix	  -> \pfx{Dr}
%% GivenName  -> \fnm{Joergen W.}
%% Particle	  -> \spfx{van der} -> surname prefix
%% FamilyName -> \sur{Ploeg}
%% Suffix     -> \sfx{IV}
%% NatureName -> \tanm{Poet Laureate} -> Title after name
%% Degrees	  -> \dgr{MSc, PhD}
%%=======================================================%%

\author*[1]{\fnm{Gijs W.A.} \sur{Simons}}\email{g.w.a.simons@tue.nl}
\author[1]{\fnm{Rik F.J.} \sur{van Haren}}
%\author[3]{\fnm{Yuqing} \sur{Jiao}}
\author[1]{\fnm{Bert} \sur{Koopmans}}

\affil[1]{\orgdiv{Department of Applied Physics and Science Education}, \orgname{Eindhoven University of Technology}, \orgaddress{\city{Eindhoven}, \postcode{5612 AZ}, \country{Netherlands}}}
%\affil[2]{\orgname{LioniX International}, \orgaddress{\city{Enschede}, \postcode{7521 AN}, \country{Netherlands}}}
%\affil[3]{\orgdiv{Department of Electrical Engineering}, \orgname{Eindhoven University of Technology}, \orgaddress{\city{Eindhoven}, \postcode{5612 AZ}, \country{Netherlands}}}

%%=======================================================%%
%% sample for unstructured abstract                      %%
%%=======================================================%%
\abstract{
    \textbf{\noindent Precise control of magnetic domain formation at the nanoscale remains constrained by stochastic defect-mediated and unstable pinning, limiting scalability and reproducibility in spintronic architectures. Here we demonstrate that spatially engineered anisotropy gradients provide a deterministic alternative. Using focused Ga$^{+}$-ion irradiation, we pattern magnetic energy landscapes containing nanoscale ``anisotropy wells'' that confine magnetic domain walls and enable bidirectional sequential switching without reliance on difficult-to-control material disorder. An analytical framework describing domain-wall energetics in graded anisotropy profiles yields predictive design rules for depinning and stability, which are supported by micromagnetic simulations and experiments. We realize programmable multi-domain configurations in continuous ferromagnetic films and demonstrate robust, reproducible switching of 750~nm regions, while first results for 100~nm are shown, approaching the theoretical limit set by the domain-wall width. By replacing unstable pinning with engineered energy landscapes, this anisotropy landscape establishes a scalable materials strategy for deterministic magnetic-state programming and opens a pathway toward dense, energy-efficient spintronic and reconfigurable magnetic nanodevices.
    }
}

\keywords{Ga$^{+}$-ion irradiation, domain-wall pinning, nanomagnetism, spintronics}

%\pacs[JEL Classification]{D8, H51}
%\pacs[MSC Classification]{35A01, 65L10, 65L12, 65L20, 65L70}

\maketitle
\newpage
%\linenumbers

\noindent Magnetic domain walls (DWs) form through the gradual reorientation of magnetic moments and define interfaces between neighboring magnetic domains, making them central to many proposed nanomagnetic information-processing schemes, including racetrack-based storage \cite{parkin_memory_2015}, neuromorphic hardware \cite{grollier_neuromorphic_2020}, and other reconfigurable magnetic architectures \cite{vidamour_reconfigurable_2023}. Although their manipulation \cite{kurian_deterministic_2023, li_ultrafast_2023} or nucleation \cite{buttner_field-free_2017, fallon_controlled_2020} can be efficiently achieved using magnetic fields or spin-torque mechanisms, realizing domain configurations that are simultaneously stable, independently addressable, and reproducibly switchable at nanometer length scales remains challenging. Established pinning approaches that exploit structural defects \cite{koyama_observation_2011} or geometric confinement \cite{bogart_dependence_2009, glathe_magnetic_2012} often lead to stochastic switching behavior, strong sensitivity to fabrication variability, and poor scalability as device dimensions are pushed toward the nanoscale.

\begin{figure*}[t]
    \centering
    \includegraphics[width=0.8\linewidth]{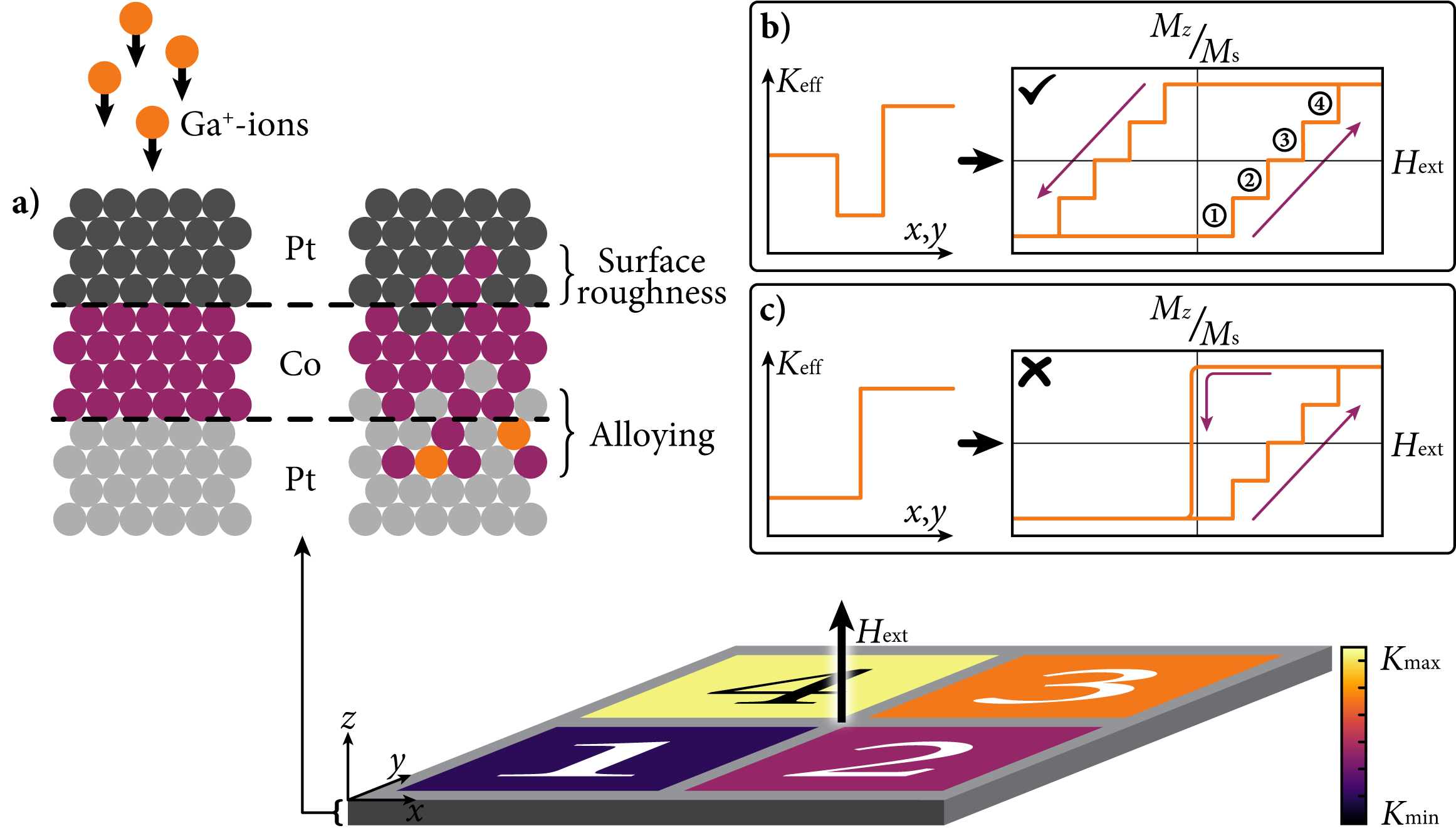}
    \caption{\textbf{a)} Schematic illustration of a focused \ce{Ga}$^{+}$-ion beam locally modifying an initially perpendicularly magnetized \ce{Pt}/\ce{Co}/\ce{Pt} stack. Ion irradiation reduces the perpendicular magnetic anisotropy $K_{\mathrm{eff}}$ in a dose-dependent manner, enabling continuous and spatially resolved tailoring of the magnetic energy landscape. \textbf{b)} Schematic anisotropy landscape across the transition between two magnetic domains with distinct anisotropy levels within the $2 \times 2$ array, separated by an engineered low-anisotropy well. The resulting stepped hysteresis behavior reflects deterministic and sequential magnetization reversal of the individual domains. \textbf{c)} Corresponding magnetic response in the absence of anisotropy wells, illustrating unstable domain configurations and no bidirectional switching behavior.}
    \label{fig: Fig1_concept}
\end{figure*}

An alternative route is to directly engineer the magnetic energy landscape through spatial modulation of the effective magnetic anisotropy. Focused ion beam (FIB) irradiation has emerged as a particularly versatile tool in this context, enabling local and continuous tailoring of magnetic properties in thin-film heterostructures with nanometer precision \cite{chappert_planar_1998, devolder_light_2000, vieu_modifications_2002, franken_precise_2011, balk_simultaneous_2017, sud_tailoring_2021}. Although other approaches such as laser annealing can also induce local magnetic property modifications \cite{ando_localised_1983, kisielewski_irreversible_2014, wu_freeform_2024, riddiford_two-dimensional_2025}, their lateral writing resolution is typically limited to the 10~$\upmu$m range, restricting nanoscale patterning.

In widely used \ce{Pt}/\ce{Co}/\ce{Pt} multilayers, the perpendicular magnetic anisotropy (PMA) originates from sharp \ce{Pt}/\ce{Co} interfaces and is highly sensitive to interfacial intermixing. Low-dose \ce{Ga}$^{+}$-ion irradiation partially disrupts these interfaces, reducing the PMA while preserving ferromagnetic order, whereas higher doses promote alloying and effectively thin the \ce{Co} layer \cite{devolder_light_2000, rettner_characterization_2002, balk_simultaneous_2017}. These dose-dependent effects, of which a schematic overview is provided in Fig.~\ref{fig: Fig1_concept}\textcolor{blue}{a}, allow the anisotropy to be tuned locally and continuously, providing a means to sculpt spatially varying magnetic energy profiles within an otherwise uniform film.

Such anisotropy engineering offers the prospect of designing DW pinning \textit{a priori}, enabling stabilization of multiple domains of arbitrary geometry and control over their switching sequence through energetic considerations rather than disorder \cite{franken_domain-wall_2011}. However, the conditions under which engineered anisotropy landscapes yield deterministic, bidirectional switching and how these conditions scale with device size and domain density remain incompletely understood. Previous studies have largely focused on multilevel behavior in micron-scale regions \cite{ma_magnetic_2025} under a bias field \cite{kurian_deterministic_2023} or isolated skyrmions \cite{sapozhnikov_artificial_2020}. This leaves the deterministic engineering of densely packed and independently addressable nanodomains with distinct remanent states within arbitrary, complex geometries without the need of a stabilizing external magnetic field an outstanding challenge.

In this paper, we introduce a general framework for deterministic domain switching based on engineered anisotropy wells: regions of reduced anisotropy relative to their surroundings. We show that “two-sided” anisotropy pinning (Fig.~\ref{fig: Fig1_concept}\textcolor{blue}{b}) enables robust DW confinement and predictable hierarchical switching, whereas single anisotropy steps fail to support stable back switching (Fig.~\ref{fig: Fig1_concept}\textcolor{blue}{c}). By spatially designing the anisotropy profile using focused \ce{Ga}$^{+}$ irradiation, we realize stable, ordered domain configurations in a static thin-film geometry, such that removing the external magnetic field at any point during its sweep leaves a well-defined and reproducible magnetic pattern.

We combine analytical modeling, micromagnetic simulations, and experiments to establish practical design rules for reliable switching in tailored anisotropy landscapes. An analytical description of the DW energy clarifies the conditions required for deterministic behavior, while simulations demonstrate stable, sequential switching in an array of sub-100~nm magnetic domains separated by a 15~nm anisotorpy well. Experimental measurements on \ce{Ga}$^{+}$-irradiated Hall crosses validate these principles, revealing reproducible switching and robust domain stabilization at 750~nm and potentially down to 100~nm, with a clear pathway toward operation at the 50~nm scale. Together, these results establish \ce{Ga}$^{+}$-based FIB anisotropy engineering as a scalable and programmable strategy towards stabilizing nanometer-scale magnetic domains at zero applied field and enabling deterministic encoding of arbitrary magnetization patterns in otherwise homogeneous thin films through controlled magnetic-field sweeps.

\section*{Bidirectional domain-wall pinning}\label{Sec: Bidirectional domain-wall pinning}
\begin{figure}[t]
    \centering
    \includegraphics[width=0.8\linewidth]{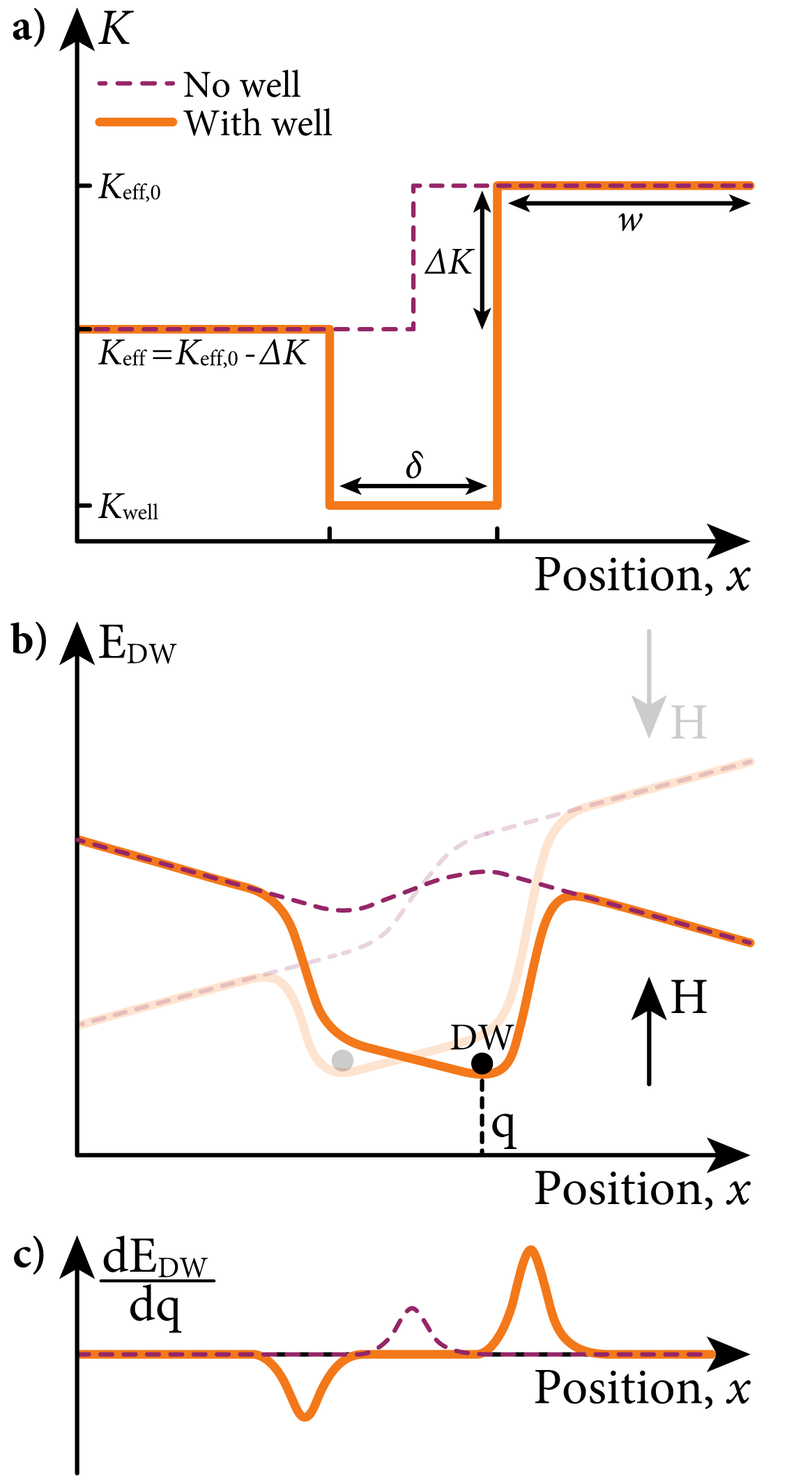}
    \caption{\textbf{a)} Schematic anisotropy profile consisting of a well with width $\delta$ and depth $K_{\mathrm{well}}$ relative to the surrounding material. \textbf{b)} Calculated energy of a Bloch-type domain wall $E_{\mathrm{DW}}$ as a function of its position $q$ under an external magnetic field $\vec{H}$. \textbf{c)} Derivative $\frac{dE_{\mathrm{DW}}}{dq}$, which determines the pinning and depinning conditions.}
    \label{fig: Fig2_AnisotropyWell_Schematic}
\end{figure}

To understand how spatial modulations of magnetic anisotropy can stabilize nanoscale magnetic domains, we developed an analytical description of a domain wall (DW) moving through an engineered anisotropy landscape. The model builds on a one-dimensional treatment \cite{franken_domain-wall_2011}, which captures Bloch-type DW pinning at abrupt changes in effective anisotropy, and we extend it to describe confinement within finite-width regions $\delta$ of reduced anisotropy $K_{\mathrm{well}}$. The essential concept is illustrated schematically in Fig.~\ref{fig: Fig2_AnisotropyWell_Schematic}: a locally defined anisotropy well generates a corresponding energy landscape for the DW determined by its derivative, whose extrema determine pinning and stability.

In perpendicularly magnetized thin films, the DW energy is primarily governed by exchange and anisotropy contributions, with external magnetic fields $\vec{H}$ providing a driving force for DW motion. Within this framework, a DW experiences an effective potential that reflects the spatial convolution of its internal magnetization profile with the local anisotropy. As a result, abrupt anisotropy changes produce sharp energy barriers and strong pinning, whereas smoothly varying anisotropy profiles yield only weak restoring forces. In the presence of an out-of-plane magnetic field, the competition between this anisotropy-derived potential and the Zeeman energy determines whether the wall remains pinned or depins.

For a single anisotropy step (purple dashed line in Fig.~\ref{fig: Fig2_AnisotropyWell_Schematic}), this framework recovers the well-established result that the depinning field is given by
\begin{equation}\label{eq: Depinning field}
    H_{\mathrm{pin}} = \frac{K_{\mathrm{eff},0} - K_{\mathrm{eff}}}{2 \mu_0 M_s},
\end{equation}
and therefore scales directly with the anisotropy contrast $\Delta K = K_{\mathrm{eff},0} - K_{\mathrm{eff}}$. Here, $K_{\mathrm{eff}}$ denotes the effective perpendicular anisotropy constant of a specific region ($K_{\mathrm{eff},0} \gt K_{\mathrm{eff}}$) and $\mu_0 M_s$ is the saturation magnetization. This contrast vanishes when the anisotropy variation occurs over a length scale much larger than the intrinsic DW width $\lambda = \sqrt{\frac{A_{\mathrm{ex}}}{K_{\mathrm{eff}}}}$, with $A_{\mathrm{ex}}$ being the exchange stiffness. While such pinning sites can impede DW motion, they do not by themselves ensure the stability of a confined magnetic domain, as the wall remains free to escape once the field direction is reversed.

To achieve \textit{bidirectional} pinning and true domain confinement, we introduce the concept of an \textit{anisotropy well}, defined as a narrow region with reduced perpendicular anisotropy $K_{\mathrm{well}}$ (tending towards the in-plane (shape) anisotropy regime, $K_{\mathrm{eff}} \lt 0$ \cite{ferre_irradiation_1999}) bounded on both sides by higher-anisotropy material:
\begin{equation}
    K\left(x\right) =
    \begin{cases}
    K_{\mathrm{eff},0} - \Delta K, & x \leq w, \\[6pt]
    K_{\mathrm{well}}, & w < x < w + \delta, \\[6pt]
    K_{\mathrm{eff},0}, & x \geq w + \delta,
    \end{cases}
\end{equation}
where $x$ denotes the position along the lateral axis, while $w$ and $\delta$ correspond to the widths of the higher-anisotropy region and the anisotropy well, respectively. The resulting anisotropy well is schematically illustrated by the solid orange profile in Fig.~\ref{fig: Fig2_AnisotropyWell_Schematic}. In this configuration, a DW entering the well encounters two opposing energy barriers located at the well interfaces. These barriers give rise to distinct forward and backward depinning fields, confining the wall within the well under both positive and negative magnetic fields. As a consequence, the enclosed magnetic domain remains stable even after the external field is removed, providing a route to zero-field domain stabilization without geometric confinement.

This analytical treatment, of which the full derivation is provided in the Supplementary Information (S.I.), relies on several simplifying assumptions, including a fixed DW profile and the neglect of thermal activation and microstructural disorder. Consequently, it captures qualitative trends and provides upper-bound estimates for depinning fields rather than quantitative predictions. Within these limitations, the model identifies the anisotropy contrast $\Delta K$ and the well width $\delta$ as the key parameters governing anisotropy-well confinement. In particular, strong bidirectional pinning requires both a sufficiently large anisotropy contrast and a well width comparable to the DW width $\lambda$. These trends establish the basis for practical anisotropy-well design, explored quantitatively in the following section using both analytical calculations and micromagnetic simulations.
% Within these limitations, the model reveals two key parameters governing the effectiveness of anisotropy wells: the anisotropy contrast $\Delta K$ and the well width $\delta$. When the well is significantly narrower than the DW width $\lambda$, the DW cannot fully relax within the reduced-anisotropy region and confinement remains weak, whereas pinning becomes strongest when the $\delta$ approaches $\lambda$, beyond which further widening yields diminishing improvement. A sufficiently large anisotropy contrast is likewise required to clearly separate the forward and backward depinning fields and to ensure robust bidirectional pinning. Together, these trends define practical design guidelines for stabilizing nanoscale magnetic domains, which we explore and validate in the following section using both the described model and numerical micromagnetic simulations.

\section*{Anisotropy well design rules}\label{Sec: Anisotropy well design rules}
Building on the analytical framework for bidirectional DW pinning, we now use the model to establish practical design rules for stable and sequentially switchable nanodomains within \ce{Pt}/\ce{Co}/\ce{Pt} trilayers. Although the analytical treatment provides primarily qualitative trends under idealized assumptions, it captures the essential interplay between DW expansion and anisotropy-induced confinement that governs stability and addressability.

\begin{figure}[t]
    \centering
    \includegraphics[width=1\linewidth]{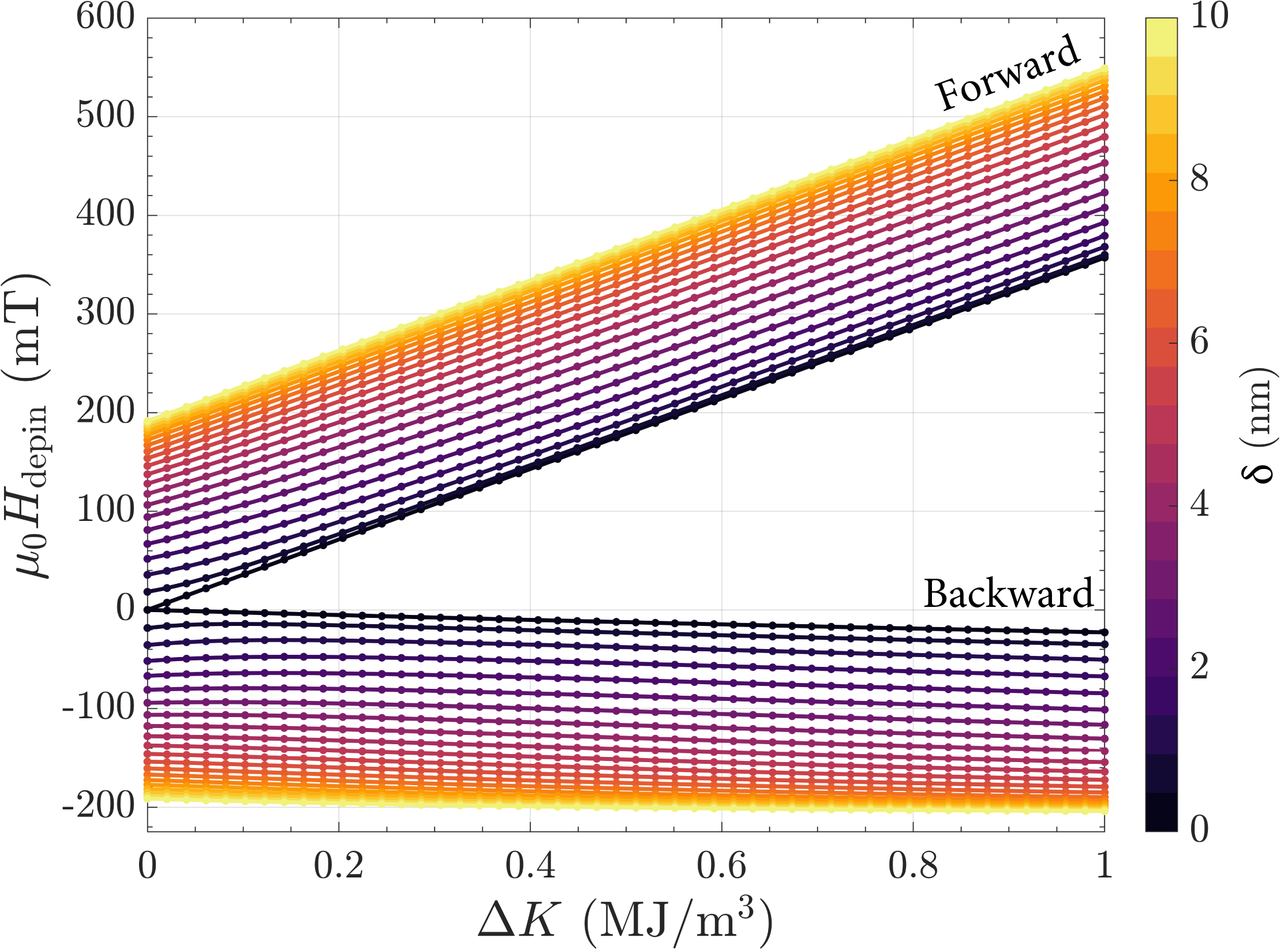}
    \caption{The calculated forward and backward depinning fields of a Bloch wall ($A_{\mathrm{ex}}=13$~pJm$^{-1}$) inside a single anisotropy well ($K_{\mathrm{well}}=-0.2$~MJ/m$^{3}$, $K_{\mathrm{eff}} = 0.5$~MJ/m$^{3}$) as a function of anisotropy contrast $\Delta K$ and well width $\delta$. The field difference between the two directions increases with $\Delta K$ and saturates as the well becomes wider than the domain-wall width.}
    \label{fig: Fig3_SingleWell_AnalyticalModel}
\end{figure}

The primary design parameters are the anisotorpy contrast $\Delta K$, the well width $\delta$, and the spacing between neighboring wells $w$. As demonstrated by Fig.~\ref{fig: Fig3_SingleWell_AnalyticalModel}, increasing $\Delta K$ enhances the asymmetry between the two interface energy barriers and thereby increases the depinning-field contrast. Meanwhile, the confinement strength depends critically on the ratio between the well width $\delta$ and the DW width $\lambda$, which determines whether the DW can fully relax within the reduced-anisotropy region. When $\delta \ll \lambda$, the DW interacts simultaneously with both interfaces, resulting in weak confinement and shallow pinning potentials. Maximum confinement is reached once $\delta \approx \lambda$, beyond which further widening yields only marginal improvement. Accordingly, the depinning-field contrast increases approximately linearly with $\Delta K = K_{\mathrm{eff}} - K_{\mathrm{well}}$ and saturates as $\delta$ approaches $\lambda$.

\begin{figure}[t]
    \centering
    \includegraphics[width=1\linewidth]{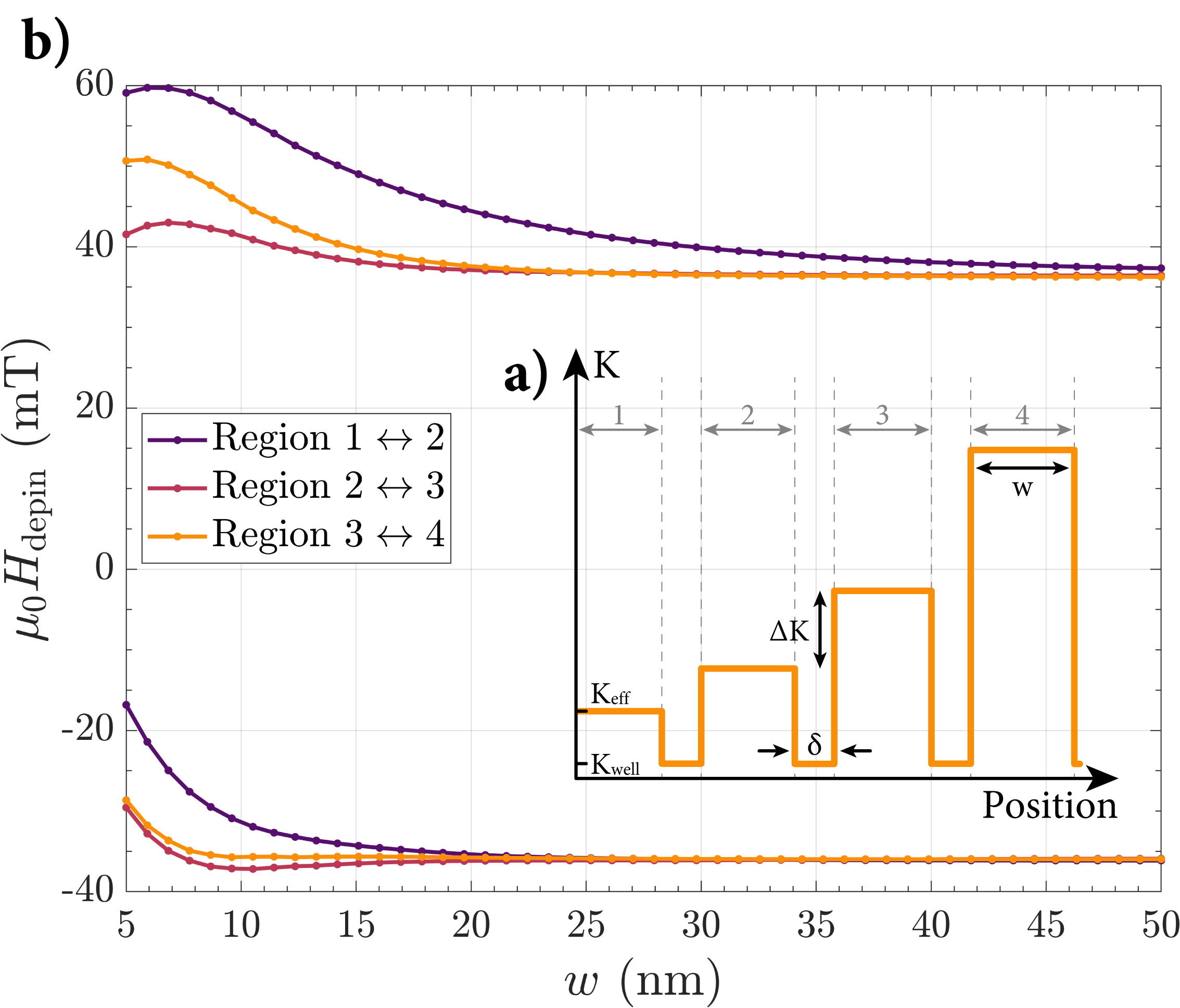}
    \caption{An example of a four-region anisotropy profile comprising three anisotropy wells ($K_{\mathrm{eff}} = 0.3$~MJm$^{-3}$, $K_{\mathrm{well}} = -0.2$~MJm$^{-3}$, $\Delta K = 0.1$~MJm$^{-3}$, and $\delta = 15$~nm). \textbf{a)} Corresponding energy landscape and \textbf{b)} depinning-field contrast between adjacent regions as a function of region width $w$. Each region supports a distinct minimum, enabling deterministic sequential switching. The contrast saturates once $w \gtrsim 50$~nm, defining the minimum size for stable, independently pinned regions.}
    \label{fig: Fig3_MultipleWells_AnalyticalModel}
\end{figure}

Extending the model to multiple wells enables evaluation of sequential switching in confined structures. Arranging wells and regions with varying magnetic anisotropy values within a finite footprint creates a cascading anisotropy landscape in which each region exhibits a distinct switching field. Fig.~\ref{fig: Fig3_MultipleWells_AnalyticalModel} shows an example of a four-region landscape calculated for $K_0 = 0.3$~MJm$^{-3}$, $K_{\mathrm{well}} = -0.2$~MJm$^{-3}$, $\Delta K = 0.1$~MJm$^{-3}$, and $\delta = 15$~nm. Deterministic switching follows when these depinning fields are sufficiently separated, such that DWs remain pinned once the external field is removed.

However, as the well width becomes smaller than the DW width ($\delta \lt \lambda$), the confined DW profile increasingly extends beyond the anisotropy barriers. While this leakage does not inherently destabilize the confined state, reducing the spacing between neighboring wells eventually causes adjacent DW profiles to interact, weakening the effective energy barriers and promoting partial coupling between neighboring domains. The model therefore predicts a practical lower size limit of approximately $w = 50$~nm for independently switchable regions in the present material system, consistent with the characteristic DW width of $\lambda \approx 10$~nm.

Thermal activation and microstructural disorder, neither included in this analytical description, further constrain the usable parameter space by reducing the effective stability of anisotropy-induced pinning barriers. Combining these considerations with the requirement of a depinning-field contrast exceeding 10~mT to account for realistic operating conditions (i.e. room temperature) defines a first-order design space for \ce{Pt}/\ce{Co}/\ce{Pt} trilayers with $\Delta K \geq 0.025$~MJm$^{-3}$, $K_{\mathrm{eff}} \geq 0.1$~MJm$^{-3}$, $K_{\mathrm{well}} < 0$~MJm$^{-3}$ (in-plane), $\delta \geq 15$~nm, and $w \geq 50$~nm. Within this regime, each anisotropy well should provide robust bidirectional confinement while minimizing interactions between neighboring regions.

The model also defines upper bounds for the number of independently switchable regions $N(W)$ with distinguishable depinning fields within a finite structure. In a strip of total width $W$, the geometric constraint is
\begin{equation}
    \begin{aligned}
        N(W) &= \frac{W}{w + \delta}, \quad \mathrm{for} \hspace{1mm} W \gg \delta, w \\
        &\leq \frac{W}{65~\mathrm{nm}},
    \end{aligned}
\end{equation}
while the tunable anisotropy range imposes
\begin{equation}
    N(K) = \frac{K_{\mathrm{max}}-K_{\mathrm{eff}}}{\Delta K},
\end{equation}
given a maximum achievable anisotropy $K_{\mathrm{max}}$. The number of addressable nanodomains is determined by the smaller of these two quantities. For a $400 \times 400$~nm$^{2}$ structure designed with the parameters above, the spatial constraint dominates and limits the system to six independently switchable regions along a single axis. These analytical bounds establish both an estimate for the minimum feature size and the maximum achievable density of sequentially addressable domains in this material system.

\section*{Micromagnetic validation and scalability}\label{Sec: Mumax3}
To assess the robustness of the analytically derived design rules under realistic magnetic interactions, we performed full micromagnetic simulations using mumax$^{3}$ at room temperature \cite{vansteenkiste_design_2014}. Unlike the one-dimensional analytical model, these simulations explicitly include dipolar interactions, the Dzyaloshinskii–Moriya interaction (DMI), and thermal fluctuations, thereby capturing the full micromagnetic energy landscape. This approach enables a rigorous assessment of whether engineered anisotropy wells can sustain deterministic and sequential switching beyond the idealized Bloch-wall approximation.

\begin{figure}[t]
    \centering
    \includegraphics[width=1\linewidth]{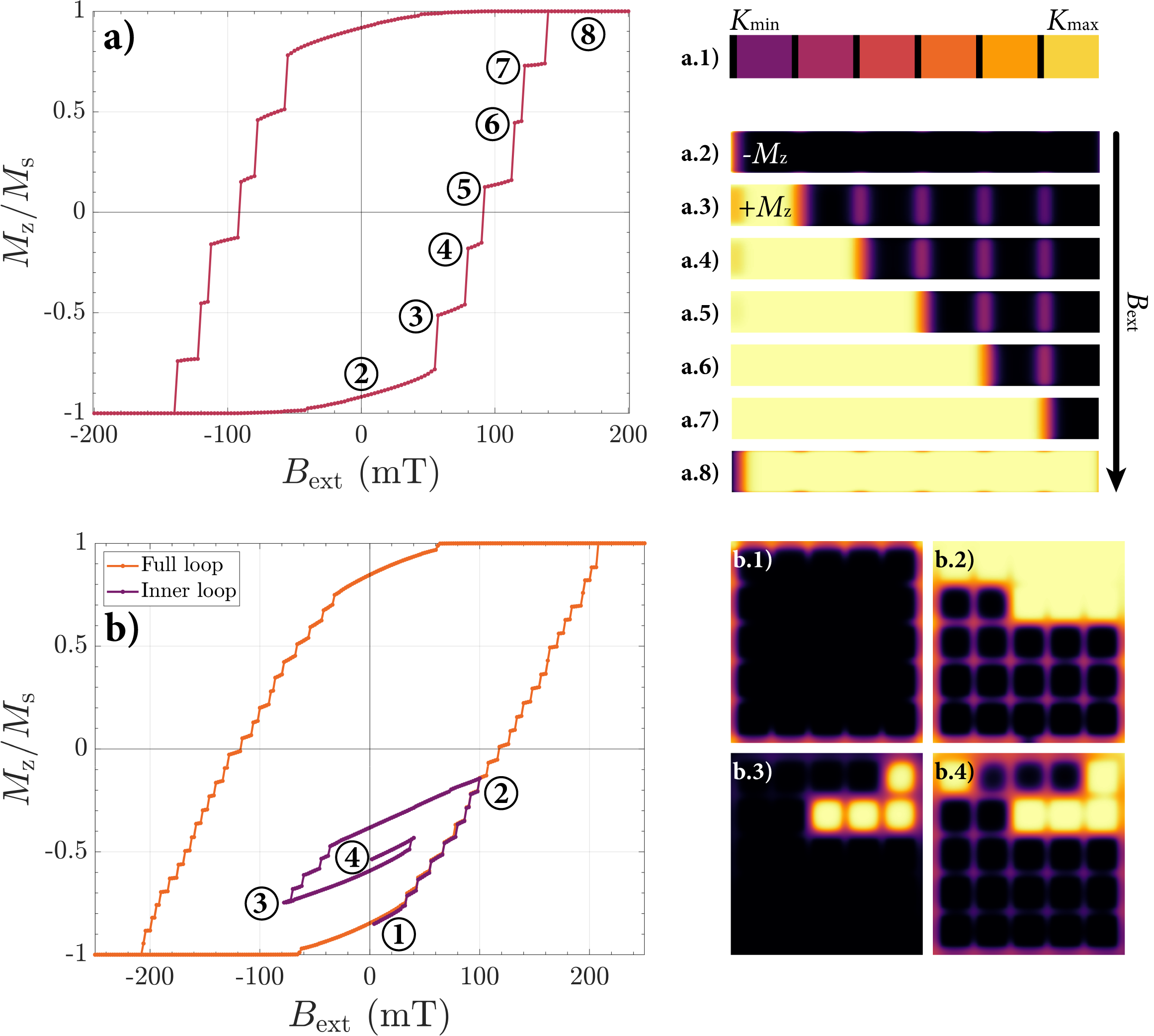}
    \caption{Micromagnetic validation of anisotropy-well–induced sequential switching. \textbf{a)} simulation of a $400 \times 400 \times 1$ nm$^{3}$ \ce{Co} film patterned into six regions separated by anisotropy wells, using $K_0 = 0.1$~MJm$^{-3}$, $K_{\mathrm{well}} = -0.2$~MJm$^{-3}$, $\Delta K = 0.03$~MJm$^{-3}$, $\delta = 15$~nm, and $w = 50$~nm. Step-like switching in the hysteresis loop reflects sequential depinning of the domain wall. \textbf{b)} Simulation of a $340 \times 340 \times 1$ nm$^{3}$ film patterned into 25 regions arranged in a grid with anisotropy values increasing in a serpentine order from 0.10 to 0.82~MJm$^{-3}$, starting from the top-left corner. The full hysteresis loop (orange) and field-pulse-induced switching sequence (purple) demonstrate the ability to program complex magnetic patterns using engineered anisotropy landscapes with $50 \times 50$~nm$^2$ magnetic elements.}
    \label{fig: Fig4_mumax3}
\end{figure}

Starting from the minimal design configuration, a $400 \times 400$~nm$^{2}$ and 1~nm thick \ce{Co} film with artificial PMA was patterned into six regions separated by anisotropy wells along a single axis (Fig.~\ref{fig: Fig4_mumax3}\textcolor{blue}{a.1}), consistent with the analytically predicted maximum for this geometry. An additional well was placed at the left boundary to emulate the initial DW assumed in the analytical model and thereby preserve symmetry in the sequential switching process. All other simulation parameters are provided in the Methods section.

The simulated hysteresis loops exhibit discrete steps corresponding to individual DW depinning events (Fig.~\ref{fig: Fig4_mumax3}\textcolor{blue}{a.3}–\textcolor{blue}{a.8}), confirming deterministic, one-by-one reversal of the patterned regions. Despite the inclusion of interactions absent from the analytical model, the switching order remains preserved. The finite hard-axis-like slope near $B_{\mathrm{ext}} = 0$~mT (Fig.~\ref{fig: Fig4_mumax3}\textcolor{blue}{a.2}) originates from the gradual rotation of magnetic moments within the anisotropy well towards the out-of-plane external field. This progressive rotation alters the magnetic environment of the switchable domains, typically leading to a reduction in the depinning field, particularly evident at higher magnetic fields in the hysteresis loops. Minor deviations in the depinning field contrast between subsequent regions, as seen in Fig.~\ref{fig: Fig4_mumax3}\textcolor{blue}{a}, indicate operation near the boundary between ideal sequential switching and partial stochasticity, largely consistent with the analytical thresholds.

% Quantitatively, the simulated depinning fields agree with analytical predictions to within $10\%$ - $20\%$. Slightly elevated switching fields in central regions arise from reduced demagnetizing energy when the DW resides near the geometric center. These effects introduce small variations in the pinning strength but do not disrupt the hierarchical switching sequence, confirming that the engineered anisotropy landscape dominates over local field inhomogeneities.

To evaluate scalability and two-dimensional addressability, we next simulated a $340 \times 340 \times 1$ nm$^{3}$ film patterned into a $5 \times 5$ grid with anisotropy values increasing in a serpentine sequence from 0.10 to 0.82~MJm$^{-3}$ (Fig.~\ref{fig: Fig4_mumax3}\textcolor{blue}{b}), corresponding to the maximum attainable $\Delta K$. Even at region sizes of approximately 50~nm, the system exhibits near-perfect sequential switching at room temperature, with 22 out of 25 regions reversing deterministically. Beyond demonstrating scalability, this cascading anisotropy landscape enables the programmable formation of complex magnetic patterns: by tailoring the spatial anisotropy distribution and sweeping the external magnetic field accordingly, predefined magnetization configurations can be written in a controlled and unambiguous manner.

These results demonstrate that two-sided anisotropy pinning remains effective in dense two-dimensional geometries and under realistic magnetic interactions. The close agreement between analytical limits and full micromagnetic simulations establishes a predictive and scalable framework for engineering high-density, programmable magnetic domain arrays.

\section*{Experimental domain switching}\label{Sec: Deterministic domain switching}
To experimentally validate the analytically and numerically predicted conditions for stable sequential switching, we implemented Ga$^{+}$-ion-induced anisotropy engineering in perpendicularly magnetized Hall-cross (HC) devices based on \ce{Ta}(2)/\ce{Pt}(2)/\ce{Co}(1)/\ce{Pt}(2) heterostructures (layer thicknesses in nm). Local irradiation sculpted the anisotropy landscapes, while magneto-optical Kerr microscopy (MOKE) and anomalous Hall effect (AHE) measurements enabled direct correlation between magnetic reversal and the engineered anisotropy profile. Device fabrication, irradiation protocols, and measurement configurations are detailed in the S.I. and Methods.

\begin{figure*}[t]
    \centering
    \includegraphics[width=0.9\linewidth]{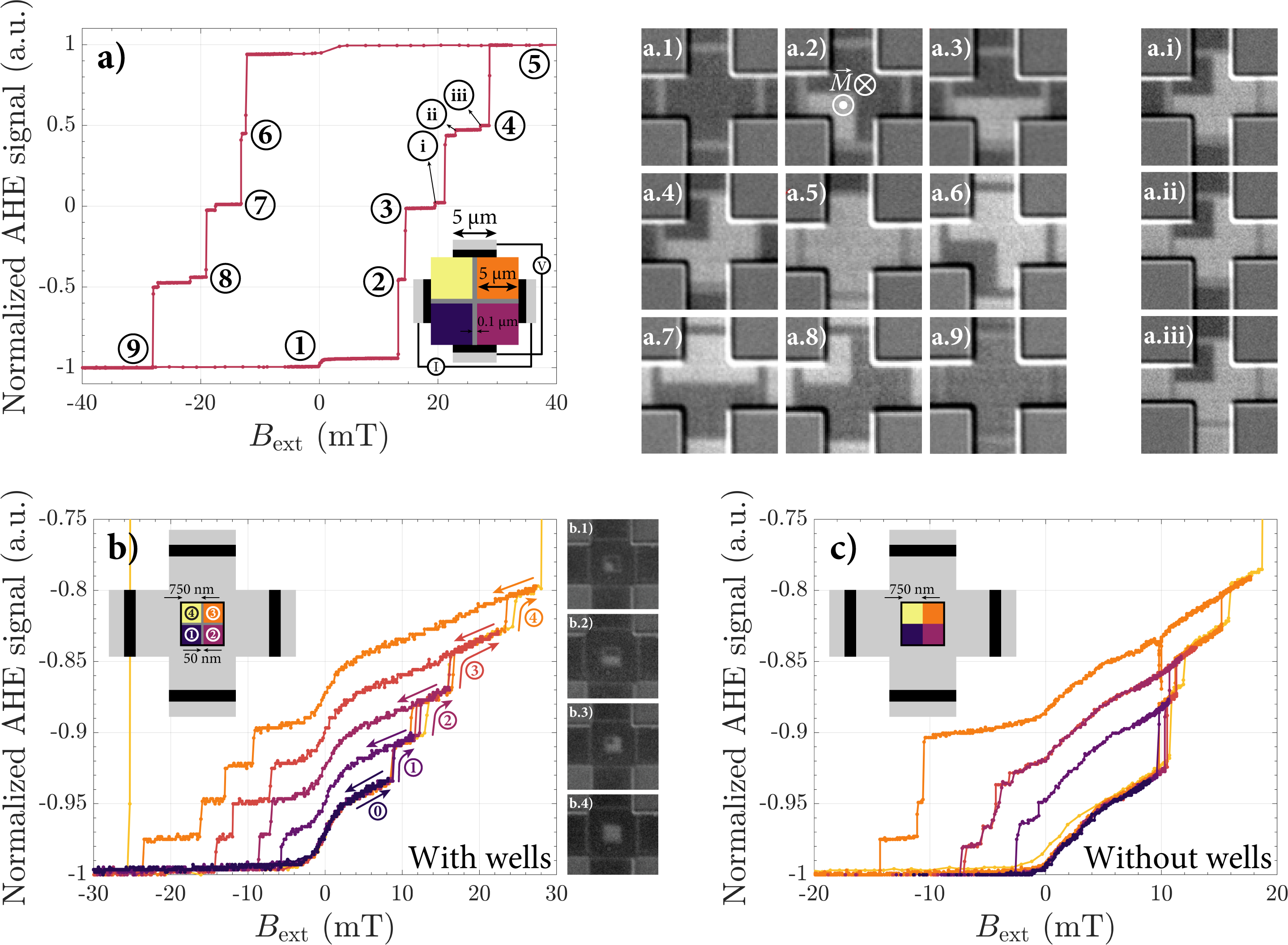}
    \caption{Experimental verification of deterministic, multi-step magnetic switching. \textbf{a)} AHE hysteresis from a micrometer-scale irradiation pattern consisting of four regions with engineered anisotropy contrast ($\Delta K = 0.05 - 0.10$~MJm$^{-3}$). Large steps (a.1-a.9) correspond to the irradiated regions as can be seen in the polar Kerr microscope images; small steps (i-iii) originate from the Hall-cross arms. \textbf{b,c)} AHE response of a submicron irradiation pattern ($w = 750$~nm, $\delta = 50$~nm) isolated by double-barrier structures to suppress arm contributions. Distinct steps arise from sequential reversal of the four regions. Hysteresis inner loops comparing \textbf{b)} patterns with wells, including corresponding Kerr microscopy images, and \textbf{c)} without wells. With wells, all intermediate states are stable and reproducible. Without wells, switching becomes non-deterministic and intermediate magnetization levels fail to repeat between loops.}
    \label{fig: Fig5_experiments}
\end{figure*}

We first designed a micrometer-scale anisotropy pattern to reproduce the predicted multi-step reversal sequence under conditions readily resolvable by Kerr microscopy. As shown in Fig.~\ref{fig: Fig5_experiments}\textcolor{blue}{a}, four regions with varied anisotropy magnitudes were written into the Hall-cross intersection with region widths of 5~$\mathrm{\upmu}$m and well widths of 100~nm. The anisotropy contrast between successive regions was set at $\Delta K = 0.05$~MJm$^{-3}$, except for one intentionally enlarged step of $\Delta K = 0.10$~MJm$^{-3}$ to test the expected increase in depinning field. The anisotropy magnitudes of the four regions, experimentally extracted via calibration (see S.I. for details), are 0.48, 0.53, 0.58, and 0.68~MJm$^{-3}$. High-dose-irradiated boundary regions (black in the figures) were introduced to magnetically isolate the patterned intersection from the Hall-cross arms, ensuring that DW depinning within the engineered anisotropy landscape precedes the propagation of DWs nucleated in the unirradiated surroundings into the sensitive cross-sectional region of the device, thereby suppressing spurious intermediate AHE levels.

Kerr microscopy (Fig.~\ref{fig: Fig5_experiments}\textcolor{blue}{a.1}-\textcolor{blue}{a.9}) reveals four distinct, sequential magnetization plateaus during field sweeps. Each region therefore acts as an independently addressable domain with a switching field defined by its engineered anisotropy. The final region, corresponding to the largest anisotropy step, requires a substantially higher reversal field, in quantitative agreement with the analytical depinning criterion.

The AHE signal exhibits additional minor steps arising from the magnetically separated Hall-cross arms, confirmed by Kerr imaging in Fig.~\ref{fig: Fig5_experiments}\textcolor{blue}{a.i}-\textcolor{blue}{a.iii}. Their relative amplitudes match the expected current-distribution weighting. Although the two regions with the highest irradiation dose occasionally reverse in close succession, consistent with their small anisotropy contrast, both MOKE and AHE measurements confirm that the overall switching hierarchy follows the designed sequence. Ga$^{+}$-induced anisotropy engineering thus enables deterministic control of multi-step magnetic reversal.

To probe scalability, we reduced the region width to $w = 750$~nm and the anisotropy-well width to $\delta = 50$~nm. Following the same magnetic isolation strategy as before, we implemented a two-level barrier design consisting of a tightly enclosing high-dose boundary around the irradiated regions and an additional outer barrier at the Hall-cross arm entrances (black in Fig.~\ref{fig: Fig5_experiments}\textcolor{blue}{b}) to suppress interference from the arms and other unintended nucleation sites. 

The full hysteresis loop in Fig.~\ref{fig: Fig5_experiments}\textcolor{blue}{b} (yellow) reveals three contributions: (1) a large step from the switching of the unirradiated surroundings between the two "black" regions occurring around $\pm 25$~mT, (2) Smaller steps corresponding to the Hall-cross arms, confirmed by Kerr images (both not visible in this zoomed view), and (3) a distinct sequence of intermediate plateaus associated with the irradiated regions. Furthermore, a distinct rounded step near zero field reflects contributions from the highly irradiated well and barrier regions, whose weak in-plane anisotropy produces a hard-axis-like offset.

To directly test whether anisotropy wells are required for stable submicron magnetic domains, we performed hysteresis inner-loop measurements on patterns with and without wells, corresponding to Fig.~\ref{fig: Fig5_experiments}\textcolor{blue}{b} and Fig.~\ref{fig: Fig5_experiments}\textcolor{blue}{c}, respectively. For patterns containing wells, restricting the field sweep to slightly beyond the switching field of each region yielded four reproducible remanent states in both sweep directions. Crucially, the zero-field magnetization levels coincide across all subloops, demonstrating that each intermediate configuration is stable, deterministic, and uniquely defined by the reversal of a single region.

In contrast, an otherwise identical pattern lacking anisotropy wells exhibits a marked loss of determinism, as observed in Fig.~\ref{fig: Fig5_experiments}\textcolor{blue}{c}. The first two regions frequently reverse together during forward sweeps, indicating insufficient pinning. More importantly, backward sweeps display irregular jumps and non-reproducible zero-field magnetization levels. These behaviors directly evidence uncontrolled DW motion and the absence of well-defined intermediate states.

These comparative measurements conclusively show that anisotropy wells are essential for achieving stable, sequential, and reproducible hierarchical switching of isolated magnetic domains. Without local minima in the anisotropy landscape, DWs do not remain confined, intermediate magnetization states are ill-defined, and deterministic behavior is lost.

\section*{Discussion}\label{Sec: Discussion}
The present results position anisotropy landscape engineering as a deterministic alternative to conventional DW pinning strategies based on structural defects or geometric constrictions. In widely studied racetrack-type concepts and nanowire devices, domain stabilization typically relies on edge roughness, lithographic notches, or material inhomogeneities that create local energy barriers \cite{koyama_observation_2011, bogart_dependence_2009, glathe_magnetic_2012}. While such approaches can generate effective pinning sites, they are stongly sensitive to fabrication variability and therefore limited scalability as feature sizes approach the DW width. In contrast, the anisotropy wells introduced in this work generate a spatially programmable energy landscape in an otherwise continuous film, enabling bidirectional confinement that is analytically predictable and experimentally reproducible. The close agreement between depinning-field contrasts, micromagnetic simulations, and measured switching sequences underscores that the dominant energy scale is the engineered anisotropy contrast rather than uncontrolled microstructural disorder.

It should be noted, however, that the present approach benefits from the use of magnetically isolated Hall-cross geometries and the selection of devices exhibiting minimal defect-induced variability. Although this enables controlled investigation of anisotropy-well-mediated confinement, disorder-driven DW pinning and stochastic propagation effects may become increasingly relevant in extended device architectures where DW transport occurs over substantially longer distances.

Beyond controlled switching, the ability to stabilize remanent intermediate states without an external magnetic field distinguishes this approach from other conventional DW systems, in which domain configurations often relax e.g. due to thermal fluctuations once the driving field or current \cite{kurian_deterministic_2023} is removed unless reinforced by structural pinning. In our approach, the two-sided anisotropy potential creates local energy minima that confine DWs and preserve uniquely defined nanodomains at zero field. This intrinsic stability enables deterministic multi-state behavior within a single continuous film, providing a mechanism to encode spatially ordered magnetic configurations without relying on geometrical segmentation. Such field-programmable and non-volatile state control is particularly relevant for emerging magnetic information technologies, including multi-level memory, reconfigurable magnonic structures, and neuromorphic hardware, where reproducibility and retention of intermediate states are essential.

The versatility of anisotropy-gradient engineering extends beyond model geometries. As a demonstration (see S.I.), a grayscale image was converted into a GDSII file and directly translated into an irradiation pattern that defines anisotropy wells along magnetic-state boundaries. This example illustrates that arbitrary digital designs can be transformed into programmable nanoscale magnetic energy landscapes.

We further explored scaling towards 100~nm regions (see S.I.), approaching the analytically predicted limit of 50~nm set by the interplay between anisotropy contrast and DW width. Although anomalous Hall effect measurements at this scale did not allow unambiguous resolution of fully deterministic switching due to reduced signal contrast, residual multi-step features and minor-loop behavior indicate that confinement persists near this regime. The primary limitation is experimental rather than fundamental: extensive high-dose irradiation required for magnetic isolation introduces a substantial hard-axis background. With optimized irradiation geometries or enhanced readout sensitivity, deterministic switching at or below 100~nm should therefore be accessible. As the analytical framework remains valid down to the exchange length and DW width, no intrinsic scaling barrier is anticipated provided sufficient anisotropy contrast can be maintained.

\section*{Conclusion}\label{Sec: Conclusion}
It was demonstrated that engineered anisotropy wells provide a deterministic mechanism for stabilizing and addressing magnetic nanodomains in continuous thin films. An analytical extension of the domain-wall pinning model established quantitative design rules for bidirectional confinement and switching hierarchy, which were validated by micromagnetic simulations and \ce{Ga}$^{+}$-ion irradiation experiments in \ce{Pt}/\ce{Co}/\ce{Pt} trilayers. These structures exhibit sequential switching and stable remanent intermediate states, whereas control patterns without anisotropy wells show stochastic and volatile behavior, confirming that programmable energy minima are essential for reliable domain confinement.

By replacing defect-mediated pinning with designed anisotropy landscapes, this approach transforms DW control from a stochastic to a programmable, analytically predictable, and two-dimensional scalable process. The resulting ability to define, stabilize, and reproducibly access dense magnetic configurations establishes a scalable framework for nanoscale magnetic state engineering and provides a foundation for spintronic devices based on deterministic DW manipulation.

\backmatter
\noindent\textbf{\large \\ Methods}\\ \vspace{-0.5cm}
\bmhead{Micromagnetic simulations}
Micromagnetic simulations were performed using the GPU-accelerated software package mumax$^{3}$ (version 3.10) \cite{vansteenkiste_design_2014} to evaluate domain stability and switching behavior in anisotropy-engineered nanostructures. The simulated system consisted of a 1-nm-thick ferromagnetic layer with lateral dimensions of approximately $340 \times 340$~nm$^{2}$, discretized into $256 \times 256 \times 1$ finite-difference cells, yielding an in-plane cell size of 1.33~nm and ensuring adequate resolution of the DW structure.

The magnetic material parameters were chosen to represent \ce{Pt}/\ce{Co}/\ce{Pt} multilayers and were set to a saturation magnetization $M_\mathrm{s} = 1.4 \cdot 10^{6}$~Am$^{-1}$, exchange stiffness $A_{\mathrm{ex}} = 13$~pJm$^{-1}$, Gilbert damping constant $\alpha = 0.02$, and temperature $T = 293$~K. A small interfacial Dzyaloshinskii–Moriya interaction (DMI) constant of $D = 0.1$~mJm$^{-2}$ was included to reflect the asymmetric interfaces present in experimental stacks. The uniaxial anisotropy axis was defined perpendicular to the film plane.

Spatially varying anisotropy landscapes were implemented by defining multiple rectangular regions with independently assigned anisotropy constants. The structure consisted of a $5 \times 5$ grid of square domains with lateral dimensions of 50~nm, separated by anisotropy wells of width 15~nm. The effective anisotropy in the domains was varied incrementally in steps of $\Delta K = 0.03$~MJm$^{-3}$ starting from a base value of $K_{\mathrm{eff}} = 0.10$~MJm$^{-3}$, while the anisotropy within the wells was set to $K_{\mathrm{well}} = -0.20$~MJm$^{-3}$ to create local energy extrema.

The magnetization was initialized in a uniform out-of-plane state and relaxed to equilibrium using the conjugate-gradient energy minimization implemented in mumax$^{3}$. Quasi-static hysteresis loops were then simulated by applying an external magnetic field perpendicular to the film plane and sweeping its magnitude in discrete steps of 2~mT. After each field step, the system was relaxed to its equilibrium state using the default energy minimization routine of mumax$^{3}$, while magnetostatic interactions were fully included via the built-in demagnetization field. Both full and minor hysteresis loops were computed, and the average magnetization and spatial distributions were recorded to extract switching fields, reversal sequence, and remanent domain configurations.

\bmhead{Hall Cross Device Fabrication}
Hall-cross devices were fabricated on Si substrates with a 100-nm thermally grown \ce{SiO2} layer. Device patterns were first defined by electron-beam lithography using a positive-tone resist, followed by DC magnetron sputter deposition of the magnetic multilayer stack \ce{Ta}(2)/\ce{Pt}(2)/\ce{Co}(1)/\ce{Pt}(2) (thicknesses in nanometers). Subsequent lift-off defined the final Hall-cross geometry. This material system was selected for its well-established perpendicular magnetic anisotropy arising from the \ce{Pt}/\ce{Co} interfaces.

Hall crosses with central intersection areas of $5 \times 5$~\textmu m$^{2}$ and $1 \times 1$~\textmu m$^{2}$ were fabricated to investigate size-dependent magnetic switching and to provide well-defined regions for localized anisotropy engineering and electrical detection via the anomalous Hall effect.

\bmhead{Experimental Setup}
Anomalous Hall effect (AHE) measurements were performed using a standard low-frequency lock-in technique. The sample was mounted on a motorized rotation stage positioned between the poles of a calibrated electromagnet providing magnetic fields up to 2~T. The stage allowed full 360$^{\circ}$ rotation, enabling application of magnetic fields at arbitrary angles relative to the sample plane, needed for the effective anisotropy extraction (as described in the S.I.). An alternating current was supplied to the Hall cross, and the transverse voltage was detected using a lock-in amplifier referenced to the current frequency.

Magnetization reversal was independently verified using a magneto-optical Kerr effect (MOKE) microscope in polar configuration, which probes the out-of-plane magnetization. Differential imaging relative to a saturated reference state was used to enhance magnetic contrast. A small out-of-plane magnetic field (up to approximately 50~mT) was applied using a calibrated copper coil. Due to coil heating and the absence of an in-situ field probe, the applied field values have a small calibration uncertainty. The optical resolution is diffraction-limited and therefore does not resolve individual nanoscale domains; however, this does not affect the interpretation of the single ultrathin ferromagnetic layer.

\bmhead{Focused Ga$^{+}$-ion beam irradiation}
A first step toward controlled domain formation was establishing the quantitative relationship between the effective perpendicular anisotropy $K_{\mathrm{eff}}$ and the Ga$^{+}$-ion dose (Full overview presented in S.I.). Focused ion irradiation was performed using a dual beam (SEM/FIB) system operated at an acceleration voltage of 30~keV and a beam current of 1~pA, corresponding to a nominal spot size of 5.3~nm. A 25\% beam overlap and a minimum dwell time of 25~ns were used to ensure uniform modification while achieving the lowest possible dose. Square regions at the center of each Hall cross were irradiated with doses starting from $1.95 \cdot 10^{-15}$~C$\upmu$m$^{-2}$ (equivalent to $0.125 \cdot 10^{13}$~ions~cm$^{-2}$), with higher doses obtained by increasing the number of beam passes.

Following irradiation, anomalous Hall effect hysteresis loops were measured under oblique magnetic fields (40$^{\circ}$ - 70$^{\circ}$ relative to the surface normal) and analyzed using a Stoner-Wohlfarth model to extract $K_{\mathrm{eff}}$. The resulting calibration curve shows a monotonic reduction of anisotropy with increasing ion dose, evolving from strong perpendicular magnetic anisotropy ($K_{\mathrm{eff}} \approx 0.8$~MJm$^{-3}$) toward near in-plane alignment at higher doses, while preserving ferromagnetic order. This dose–anisotropy relationship was subsequently used to define spatially varying anisotropy landscapes via standard GDSII pattern files and an automated irradiation routine, enabling deterministic programming of arbitrary magnetic energy profiles within continuous ferromagnetic films. After irradiation, devices were wire-bonded using ultrasonic gold bonding to enable electrical measurements.

\bmhead{Supplementary information}
All supplementary data is provided at the end of this manuscript or as a separate document. Further supporting data might be available upon reasonable request.

\bmhead{Contribution}
G.W.A. Simons conceived and designed the project. G.W.A. Simons and R.F.J. van Haren were responsible for the fabrication of the devices. R.F.J. van Haren was responsible for the machine automization. R.F.J. van Haren and G.W.A. Simons developed the experimental protocol, performed simulations, and collected data. G.W.A. Simons and B. Koopmans supervised the project. G.W.A. Simons wrote the original manuscript draft. All authors contributed to writing and editing the manuscript.

\bmhead{Acknowledgments}
This project received funding from the PhotonDelta National Growth Fund. For the purpose of open access, a CC BY public copyright license is applied to any Author Accepted Manuscript version arising from this submission.

\bmhead{Competing Interests}
The authors declare no conflicts of interest.
%\section*{Declarations}

%\newpage
\bibliographystyle{ieeetr}
\bibliography{sn-bibliography}

%% if required, the content of bbl file can be included here once bbl is generated
%%\input sn-articlebbl

%\input sn-sample-bib.tex%

\clearpage

\begin{appendices}
\section{Analytical model formulation}\label{section: Analytical model derivation}
To model domain-wall (DW) pinning in spatially varying anisotropy landscapes, we consider a one-dimensional Bloch-type DW in a perpendicularly magnetized thin film like \ce{Pt}/\ce{Co}/\ce{Pt}. This treatment assumes translational invariance along the wall plane. Such a one-dimensional approximation is appropriate for extended films and micron-scale conduits where the DW does not experience any micromagnetic changes along its own width, but it becomes less exact in laterally confined geometries such as Hall crosses, where two-dimensional effects can modify the detailed pinning fields.

Within this approximation, the equilibrium magnetization profile minimizing the micromagnetic energy and determined by the angle between neighboring spins is given by 
\begin{equation}\label{eq: Bloch profile}
    \theta \left( x \right) = \pm 2 \arctan \left( \exp \left[ \frac{x-q}{\lambda} \right] \right),
\end{equation}
where $q$ denotes the DW position and $\lambda=\sqrt{A_{\mathrm{ex}}/K_{\mathrm{eff}}}$ is the characteristic wall width determined by the exchange stiffness $A_{\mathrm{ex}}$ and the effective anisotropy $K_{\mathrm{eff}}$. The azimuthal angle between neighboring spins is fixed at $\phi = \pm \pi/2$ corresponding to a Bloch wall.

Although interfacial Dzyaloshinskii–Moriya interaction (DMI) is expected in \ce{Pt}/\ce{Co}/\ce{Pt} multilayers, its magnitude remains small compared with the effective anisotropy ($D^2/4A_{\mathrm{ex}} \lt K_{\mathrm{eff}}$), such that it primarily modifies wall chirality rather than the energetic stability or pinning thresholds considered here. The Bloch-wall approximation therefore provides an accurate description of the relevant energetics.

Following Franken \textit{et al.} for a single anisotropy step \cite{franken_domain-wall_2011}, and assuming that $\lambda$ remains approximately constant over the region of interest, the DW energy per unit area in the presence of a spatially varying anisotropy $K\left(x\right)$ and centered around $q$ can be written as
\begin{equation}\label{eq: DW energy}
    E_{\mathrm{DW}}\left(q\right) = \int_{-\infty}^{\infty} K\left(x\right) \sech^2 \left( \frac{x-q}{\lambda} \right),
\end{equation}
which is solely based on exchange contributions and the effective anisotropy. This expression shows that the wall experiences an effective energy landscape defined by the convolution of the anisotropy profile with the intrinsic wall shape. The assumption of constant $\lambda$ simplifies the analysis but constitutes an approximation, since in general $\lambda \left( x \right)=\sqrt{A_{\mathrm{ex}}/K\left( x \right)}$ varies with position. For moderate anisotropy variations, this dependence produces only small corrections, and the constant-width approximation accurately captures the scaling of pinning energies. However, for strong anisotorpy gradients, the true wall structure deviates slightly form the assumed profile, limiting the quantitative accuracy of the model.

In the presence of an external out-of-plane magnetic field $H$, the total energy becomes
\begin{equation}\label{eq: Total energy}
    E\left(q\right) = E_{\mathrm{DW}}\left(q\right) - 2 \mu_0 M_s H q,
\end{equation}
where the second term represents the Zeeman contribution and $\mu_0 M_s$ is the saturation magnetization. Here, the depinning field $H_{\mathrm{pin}}$ corresponds to the maximum restoring force exerted by the anisotropy landscape,
\begin{equation}\label{eq: Max slope}
    \max _{-\infty<q<\infty} \left\lvert \frac{dE_{\mathrm{DW}}}{dq} \right\rvert = 2 \mu_0 M_s H_{\mathrm{pin}}.
\end{equation}
For an abrupt anisotropy step with a transition width much smaller than the DW width ($a \ll \lambda$), this reduces to
\begin{equation}\label{eq: Depinning field}
    \lim _{a \rightarrow 0} H_{\mathrm{pin}} = \frac{K_{\mathrm{eff},0} - K_{\mathrm{eff}}}{2 \mu_0 M_s},
\end{equation}
whereas for smooth variations, $a \gg \lambda$, the pinning field approaches zero, reflecting the reduced energy gradient experienced by the DW.

Bidirectional pinning is obtained by introducing an anisotropy well consisting of a reduced-anisotropy region of width $\delta$ bounded by higher-anisotropy regions,
\begin{equation}
    K\left(x\right) =
    \begin{cases}
    K_{\mathrm{eff},0} - \Delta K, & x \leq w, \\[6pt]
    K_{\mathrm{well}}, & w < x < w + \delta, \\[6pt]
    K_{\mathrm{eff},0}, & x \geq w + \delta,
    \end{cases}
    \label{Appx eq: anisotropy well profile}
\end{equation}
Evaluating Eq.~(\ref{eq: DW energy}) for this profile yields two local energy extrema at the well boundaries, corresponding to forward and backward depinning thresholds, thereby confining the DW under small positive and negative fields. To minimize the inaccuracy of a constant DW width over such an anisotropy landscape while retaining analytical tractability, the wall width is approximated as a piecewise constant,
\begin{equation}
    \lambda= \begin{cases}\sqrt{\frac{A_{\mathrm{ex}}}{K_i}}, & K_i>0, \\ \delta + \frac{\sqrt{\frac{A_{\mathrm{ex}}}{K_{i-1}}}+\sqrt{\frac{A_{\mathrm{ex}}}{K_{i+1}}}}{2}, & K_i \leq 0,\end{cases}
\end{equation}
where $K_i$ refers to the well anisotropy and $i \pm 1$ to the regions left and right of the well, respectively. This description allows the DW to extend partially beyond the well boundaries. For strongly reduced or negative well anisotropy, the assumed continuous Bloch profile becomes less accurate, and the wall may split into two coupled $90^{\circ}$ rotations separated by an in-plane magnetized region. This configuration modifies the detailed wall structure but preserves the overall energy hierarchy and pinning asymmetry.

Thermal activation and microstructural disorder are not included in this model. In real materials, thermal fluctuations and defects assist depinning, reducing coercive and pinning fields relative to the intrinsic values calculated here. The analytical predictions therefore represent upper-bound estimates, with experimental depinning fields typically reduced by approximately one order of magnitude. Despite these limitations, the model captures the essential scaling relations linking the anisotropy parameters ($K_{\mathrm{eff}}$, $\Delta K$, $K_{\mathrm{well}}$, and $\delta$) to the resulting DW stability and switching hierarchy. These relations provide the theoretical foundation for the design rules developed in the main text.

\section{Local determination of the effective anisotropy}
To quantify the effective perpendicular anisotropy locally within irradiated regions, we employed anomalous Hall effect (AHE) measurements on lithographically defined Hall cross devices. In contrast to SQUID-VSM measurements, which probe the total magnetic moment of an entire sample, the AHE signal originates predominantly from the central intersection of a Hall cross device, enabling spatially selective characterization of the modified region. When a current is applied along one arm of the cross, the transverse voltage is proportional to the out-of-plane component of the magnetization $M_z$ through the anomalous Hall effect, thereby providing a direct electrical probe of the local magnetic state.

\begin{figure}[b]
    \centering
    \includegraphics[width=0.9\linewidth]{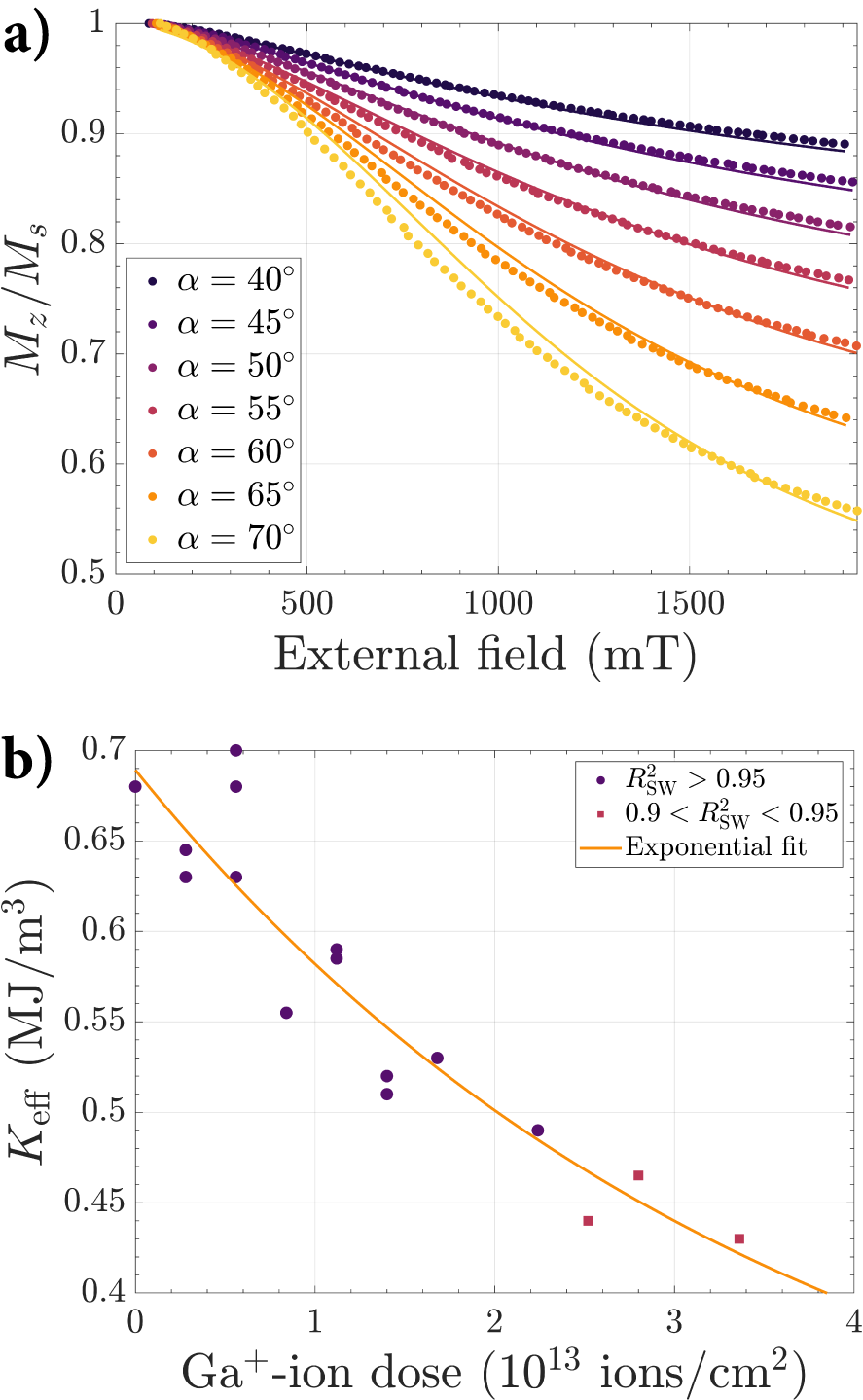}
    \caption{\textbf{a)} Segments of anomalous Hall effect hysteresis loops measured on a \ce{Pt}(2 nm)/\ce{Co}(1 nm)/\ce{Pt}(2 nm) Hall cross for different angles between the applied magnetic field and the surface normal. Solid lines show fits based on the Stoner–Wohlfarth model, from which the effective perpendicular anisotropy $K_{\mathrm{eff}}$ is extracted \textbf{b)} Calibration of the effective anisotropy as a function of \ce{Ga}$^{+}$-ion dose. Increasing irradiation progressively reduces $K_{\mathrm{eff}}$, driving the system toward in-plane anisotropy and reducing the agreement with the Stoner–Wohlfarth model, as reflected by the coefficient of determination $R_{\mathrm{SW}}^2$.}
    \label{fig: FigAppx1_SWfit}
\end{figure}

The effective anisotropy was extracted by measuring AHE hysteresis loops while applying an external magnetic field at oblique angles $\alpha$ relative to the film normal, as can be seen in Fig.~\ref{fig: FigAppx1_SWfit}\textcolor{blue}{a}. The field angle was varied between $40^{\circ}$ and $75^{\circ}$ in steps of $5^{\circ}$, and the field magnitude was swept between -2~T and 2~T. Tilting the magnetic field progressively rotates the magnetization away from the perpendicular easy axis toward the in-plane hard axis. The field-dependent evolution of $M_z$ reflects the competition between anisotropy and Zeeman energies, with stronger perpendicular anisotropy requiring larger fields to achieve comparable rotation.

The experimental data were analyzed using the Stoner–Wohlfarth model, which describes the magnetization as a single uniform vector. Although this macrospin approximation does not capture domain formation or DW motion, it provides a reliable estimate of the average effective anisotropy in continuous thin films. Within this model, the magnetic energy density is given by
\begin{equation}
    \varepsilon_{\mathrm{SW}}=K_{\mathrm{eff}} \sin ^2(\theta)-\mu_0 H M_{\mathrm{s}} \cos (\alpha-\theta),
\end{equation}
where $\theta$ is the angle between the magnetization and the surface normal, $K_{\mathrm{eff}}$ is the effective anisotropy, $M_s$ is the saturation magnetization, and $H$ is the applied magnetic field. For each field value and angle, the equilibrium magnetization direction was obtained by numerical minimization of this energy. The resulting magnetization projection was then fitted to the normalized AHE signal, visualized in Fig.~\ref{fig: FigAppx1_SWfit}\textcolor{blue}{a} by the solid lines.

Prior to fitting, a linear background due to the ordinary Hall effect was subtracted and the signal was normalized to the saturation value. The effective anisotropy was determined through an iterative fitting procedure that maximized the coefficient of determination $R^2$ between model and experiment. Fits with $R^2 \gt 0.9$ were considered acceptable, indicating good agreement with the macrospin model.

Because the Stoner–Wohlfarth model neglects thermal activation, spatial non-uniformities, and domain formation, the extracted values represent an effective anisotropy averaged over the probed region rather than an exact intrinsic material constant. Nevertheless, this approach provides a robust and widely used method to quantify local anisotropy variations in patterned magnetic nanostructures and forms the basis for the irradiation-dependent anisotropy calibration used in this work.

Fig.~\ref{fig: FigAppx1_SWfit}\textcolor{blue}{b} shows the resulting calibration curve of the effective perpendicular anisotropy $K_{\mathrm{eff}}$ as a function of \ce{Ga}$^{+}$-ion dose. The data are grouped according to the fit quality, quantified by $R^2$. As expected, increasing ion dose leads to a continuous reduction of $K_{\mathrm{eff}}$, progressively driving the system toward in-plane anisotropy. Concurrently, the agreement with the Stoner–Wohlfarth model deteriorates, reflecting the breakdown of the single-domain approximation as the perpendicular anisotropy weakens. This behavior establishes an upper practical dose limit for reliable and predictable anisotropy engineering.

\section{Measurements of 100~nm magnetic domains}
Having established deterministic switching and zero-field stability in larger structures, we next explored the scalability of anisotropy-engineered confinement toward the predicted theoretical limit of approximately 50~nm. This required improved patterning accuracy, enabling reliable irradiation of 1~$\upmu$m Hall crosses containing regions as small as 100~nm. To isolate the switching signal of the nanodomains, a modified irradiation geometry was implemented in which the Hall cross arms and surrounding areas were irradiated at high doses to suppress their perpendicular anisotropy. The four~100 nm regions were defined using a base dose of $0.125 \cdot 10^{13}$ ions cm$^{-1}$, with the ion dose in successive regions increased in integer multiples of this value to create a controlled switching hierarchy through progressively reduced effective anisotropy, while the separating wells were maintained at a width of 50~nm to preserve bidirectional pinning, consistent with the parameter space used for the larger magnetic regions in Fig.~\ref{fig: Fig5_experiments}\textcolor{blue}{b}.

\begin{figure}[t]
    \centering
    \includegraphics[width=0.9\linewidth]{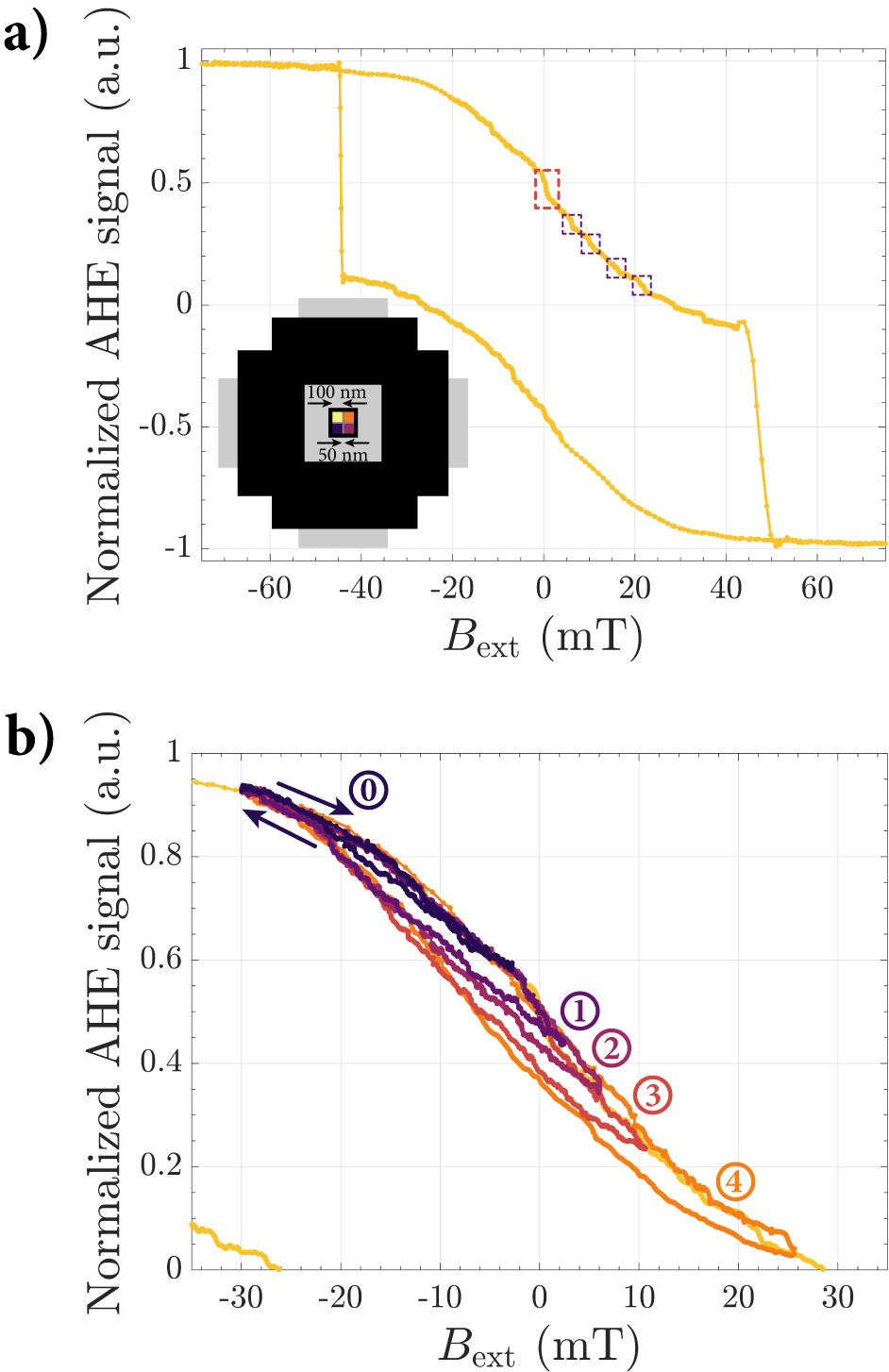}
    \caption{\textbf{a)} A out-of-plane (OOP) magnetization ($M_z/M_s$) hysteresis loop measured via the anomalous Hall effect during a full field sweep of a $1 \times 1$~$\upmu$m$^2$ Hall cross with four $100 \times 100$ nm$^{2}$ distinct magnetic regions. The red dashed box indicates the signal attributed to the canting of the 50~nm anisotropy well magnetization towards the OOP external magnetic field, which serves as an internal amplitude reference. Dark purple dashed boxes mark features tentatively assigned to switching of individual 100~nm regions \textbf{b)} Inner hysteresis loops measured around the suspected transition fields. Numbers denote the proposed number of switched regions at the termination points of the inner loops (colour coded). All data was smoothed using a fast Fourer transform filter for clarity.}
    \label{fig: FigAppx2_100nm}
\end{figure}

This approach eliminated spurious switching contributions from the Hall cross arms, resulting in a simplified hysteresis response. However, the extensive high-dose irradiation introduced a substantial hard-axis background signal that significantly reduced the relative contrast of the 100~nm regions, as seen in Fig.~\ref{fig: FigAppx2_100nm}\textcolor{blue}{a}. To identify potential switching events, the signal associated with the 50~nm anisotropy wells was used as an internal amplitude reference. The wells produced a characteristic change of approximately 0.1 $M_z$/$M_s$, highlighted by a red box, implying an expected signal of approximately 0.05 $M_z$/$M_s$ for individual $100 \times 100$ nm$^{2}$ regions. Several features of this magnitude could be tentatively identified in the anomalous Hall loop, indicated by the dark purple boxes in the figure.

To further assess their origin, a series of minor hysteresis loops was measured around the suspected transition fields and plotted in Fig.~\ref{fig: FigAppx2_100nm}\textcolor{blue}{b}. These measurements, although challenging to interpret, revealed discrete changes in remanent magnetization consistent with partial sequential switching. In particular, distinct remanent states corresponding to different numbers of reversed regions could be stabilized, and their reversal fields could be approximately identified from the convergence and separation of inner loops. However, the signal variations associated with individual switching events were comparable to the background fluctuations, and the precise switching order could not be determined unambiguously.

These measurements suggest that anisotropy-engineered confinement remains effective even at dimensions approaching 100~nm, and they provide preliminary evidence that sequential switching behavior may persist near the predicted size limit. At the same time, the reduced signal-to-background ratio highlights a key experimental challenge: large highly irradiated areas introduce hard-axis contributions that obscure nanoscale switching signals. Further optimization of the irradiation geometry to minimize such background contributions will be essential to conclusively establish deterministic switching at the smallest length scales.

\section{Complex magnetic textures}
\begin{figure*}[t]
    \centering
    \includegraphics[width=0.9\linewidth]{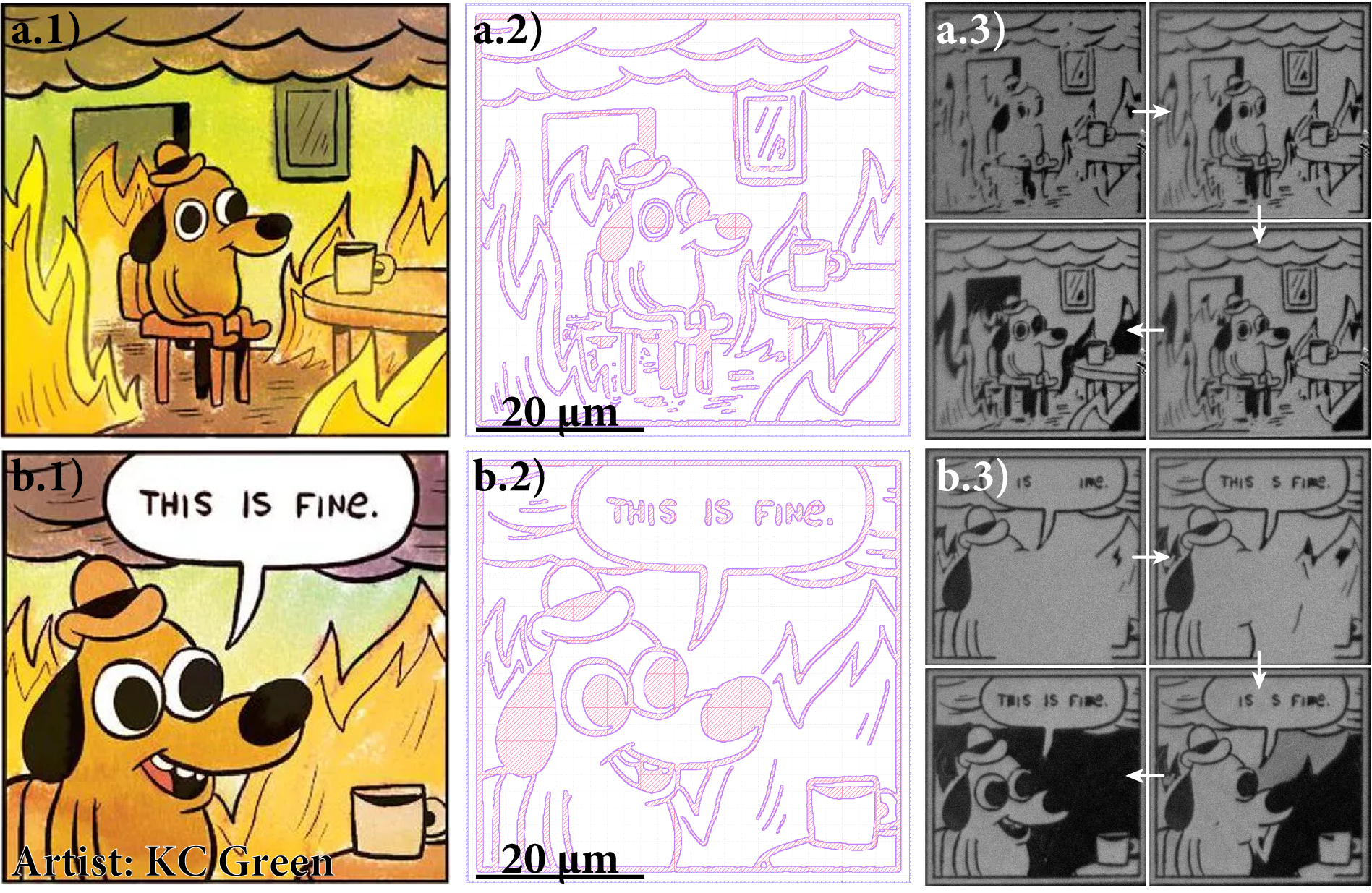}
    \caption{\textbf{a.1, b.1)} Input PNG images used as digital templates. \textbf{a.2, b.2)} Corresponding GDSII layouts generated from the images, defining anisotropy-well placement (purple) along boundaries between adjacent up- (white) and down-magnetized (red) regions. \textbf{a.3, b.3)} Polar Kerr microscopy images of the irradiated films for increasing external magnetic field, revealing the resulting out-of-plane magnetic configurations resembling the programmed patterns. Together, the panels illustrate the automated workflow from arbitrary digital image to experimentally realized magnetic energy landscape.}
    \label{fig: FigAppx3_Meme}
\end{figure*}

As a demonstrative exercise, we explored the extent to which anisotropy engineering can reproduce arbitrary magnetic landscapes. A PNG image was converted into a GDSII layout and processed by an automated routine that assigns anisotropy wells along each boundary between would-be neighboring up- and down-magnetized regions. The complete workflow is illustrated in Fig.~\ref{fig: FigAppx3_Meme}.

The resulting GDSII layout (Fig.~\ref{fig: FigAppx3_Meme}\textcolor{blue}{a.2} and \textcolor{blue}{b.2}) comprises four distinct irradiation regions. A surrounding boundary frame (blue) was exposed to a high ion dose sufficient to induce in-plane anisotropy, thereby preventing external domain walls from entering the patterned area. The white regions correspond to the intrinsic, non-irradiated perpendicular anisotropy of the magnetic film. The red regions were irradiated with a dose of $3.5 \cdot 10^{13}$~ions~cm$^{-2}$, yielding an effective anisotropy of approximately $K_{\mathrm{eff}} \approx 0.425$~MJm$^{-3}$ for the \ce{Ta}(2~nm)/\ce{Pt}(2~nm)/\ce{Co}(1~nm)/\ce{Pt}(2~nm) stack. The white and red areas are separated by anisotropy wells (purple), defined by at least twice the ion dose of the red regions to ensure strong bidirectional DW pinning.

The pattern was written directly into the continuous ferromagnetic film using focused Ga$^{+}$-ion irradiation, without manual feature-by-feature optimization. Although conceived primarily as a proof-of-principle demonstration (and admittedly for some enjoyment), this example illustrates the design freedom of the approach: complex, digitally defined magnetic energy landscapes can be translated into physical anisotropy profiles. Such flexibility highlights the potential of anisotropy patterning as a rapid prototyping tool for unconventional and application-specific magnetic architectures.

\end{appendices}

\end{document}